\documentclass[12pt]{article}   
\usepackage{psfig,times,fancyheadings}  % 17 dec. 2001: pbs avec times
\begin{document}
\thispagestyle{empty}

 \lhead[\fancyplain{}{\sl }]{\fancyplain{}{\sl }}
 \rhead[\fancyplain{}{\sl }]{\fancyplain{}{\sl }}

%%%%%%%% Pour changer les valeurs par defaut pour taille figure,
%%%%%%%% sinon au-dela d'une hauteur de 134 mm = 70% on est rejete a la fin
 \renewcommand{\topfraction}{.99}      
 \renewcommand{\bottomfraction}{.99} 
 \renewcommand{\textfraction}{.0}

%%%%% Definitions

\newcommand{\nc}{\newcommand}

\nc{\qI}[1]{\section{{#1}}}
\nc{\qA}[1]{\subsection{{#1}}}
\nc{\qun}[1]{\subsubsection{{#1}}}
\nc{\qa}[1]{\paragraph{{#1}}}

            % Enumerations
\def\qbu{\hfill \par \hskip 6mm $ \bullet $ \hskip 2mm}
\def\qee#1{\hfill \par \hskip 6mm #1 \hskip 2 mm}

\nc{\qfoot}[1]{\footnote{{#1}}}
\def\qL{\hfill \break}
\def\qpar{\vskip 2mm plus 0.2mm minus 0.2mm}
\def\qtvi{\vrule height 2pt depth 5pt width 0pt}
\def\qth{\vrule height 12pt depth 0pt width 0pt}
\def\qtb{\vrule height 0pt depth 5pt width 0pt}
\def\tvi{\vrule height 12pt depth 5pt width 0pt}

\def\qparr{ \vskip 1.0mm plus 0.2mm minus 0.2mm \hangindent=10mm
\hangafter=1}

                % Decale UN paragraphe
                % Attention! La double accolade est vitale, sinon tout le
                % est decale (cf TEX p.199)
                % On peut aller a la ligne avec \qL=\hfill \break
                % Par contre ne supporte pas les lignes blanches
\def\qdec#1{\par {\leftskip=2cm {#1} \par}}

   %% Defs specifiques
\def\qdpt{\partial_t}
\def\qdpx{\partial_x}
\def\qddpt{\partial^{2}_{t^2}}
\def\qddpx{\partial^{2}_{x^2}}
\def\qn#1{\eqno \hbox{(#1)}}
\def\qds{\displaystyle}
\def\qw{\widetilde}
\def\qmax{\mathop{\rm Max}}   % Petit livre Tex (p.167)
\def\qmin{\mathop{\rm Min}}   % Petit livre Tex (p.167)

%%%%% End of definitions

\def\qci#1{\parindent=0mm \par \small \parshape=1 1cm 15cm  #1 \par
               \normalsize}

\null
\vskip 1.5 cm

\centerline{\bf \Large Stock markets are not what we think they are:}
\vskip 5mm
\centerline{\bf \Large the key roles of cross-ownership and
corporate treasury stock}                                      

\vskip 1cm
\centerline{\bf Bertrand M. Roehner $ ^1 $ }
\vskip 4mm
         
\centerline{\bf Institute for Theoretical and High Energy Physics}
\centerline{\bf University Paris 7 }

\vskip 2cm

{\bf Abstract}\quad We describe and document three mechanisms by
which corporations can influence or even control stock prices.
(i) Parent and holding companies wield control over other publicly
traded companies. (ii) Through clever management of treasury
stock based on buyback programs and stock issuance, stock price
fluctuations can be amplified or curbed. (iii) Finally, history shows
a close interdependance between the
level of stock prices on the one hand and 
merger and acquisition activity on the other hand. 
This perspective
in which Boards of Directors of major companies shepherd the market
offers a natural interpretation of the so-called "herd behavior" observed
in stock markets. The traditional view holds that 
by driving profit expectations, corporations have an indirect role in
shaping the market. In this paper, we suggest that over the last decades
they became more and more the direct moving force of stock markets.

\vskip 1cm

\centerline{June 27, 2004}

\vskip 8mm
\centerline{\it Preliminary version, comments are welcome}

\vskip 1cm
Key-words: stock markets, cross-ownership, stock buybacks, 
corporate gouvernance, collective behavior, herd effect
\vskip 1cm 

1: ROEHNER@LPTHE.JUSSIEU.FR
\qL
\phantom{1: }Postal address where correspondence should be sent:
\qL
\phantom{1: }B. Roehner, LPTHE, University Paris 7, 2 place Jussieu, 
F-75005 Paris, France.
\qL
\phantom{1: }E-mail: roehner@lpthe.jussieu.fr
\qL
\phantom{1: }FAX: 33 1 44 27 79 90

\vfill \eject

\qI{Cross-ownership}

As everyone knows, Bill Gates owns about 10\% of the shares of
Microsoft and Jeff Bezos, the founder of Amazon holds about
25\% of its shares. Altria, the company formerly known as Philip Morris,
owns 84\% of the shares of Kraft Foods (NYSE: KFT), the US number 
one food company. In 2004
Altria's chief executive officer, Louis Camilleri, was also chairman
of Kraft. Metlife, a huge insurance company, owns 53\% of the Reinsurance
Group of America (NYSE: RGA) who is itself a multi-billion company. 
Renault, the French automaker, owns 45\% of Nissan Motors, Japan's
number two automaker after Toyoto. One may think that these cases
are exceptional. In fact, they are rather the rule. Consider the case
of Germany, one of the countries for which one has detailed 
ownership statistics. In 58\% of the publicly traded companies, the
largest stockholder has more than 50\% of the total voting stock;
in 82\% of them the largest stockholder has more than 25\% of the
voting stock (Prigge 1998).
If one remembers that a corporation $ A $ is considered 
to have {\it de jure} control of a company $ B $ if it owns at least 25\% 
of its voting shares, one sees that our previous statement was hardly
exaggerated. In the case of the corporations which make up the
Dow Jones Industrial index, table 1 shows that the percentage of
the shares which are not in the hands of  the company, of insiders
or of institutions is on average equal to 28\%.
\qpar

What are the consequences in terms of stock price behavior?
Fig. 1 shows the correlation between Renault and Nissan stock prices
over the last decade. There is a clear transition
from the period before takeover
during which the two stocks moved almost independently from each
other to a regime characterized by a cross-correlation of the order 
of 0.75. When one
party holds almost 50\% of the shares it can control the price
to some extent.
Perhaps, the majority stockholder will not care to exercise that 
control is everyday transactions, but in case of sudden price fluctuations
there is documented evidence that it will (see below).
This observation is by no means
in contradiction with the market efficiency tenet of economic theory.
As a matter of fact, the smaller the number of major players, the 
quicker the market is able to respond to new information. 
For commodities a market where one trader has about 50\% of the
inventory would hardly be considered as a perfectly competitive market
but rather as one with monopolistic competition. As a result, 
it would highly
questionable to model and simulate such a market as 
a set of numerous agents, each of whom has only
a small action on the market. 
%%%% Fig.1
  \begin{figure}[htb]
    \centerline{\psfig{width=15cm,figure=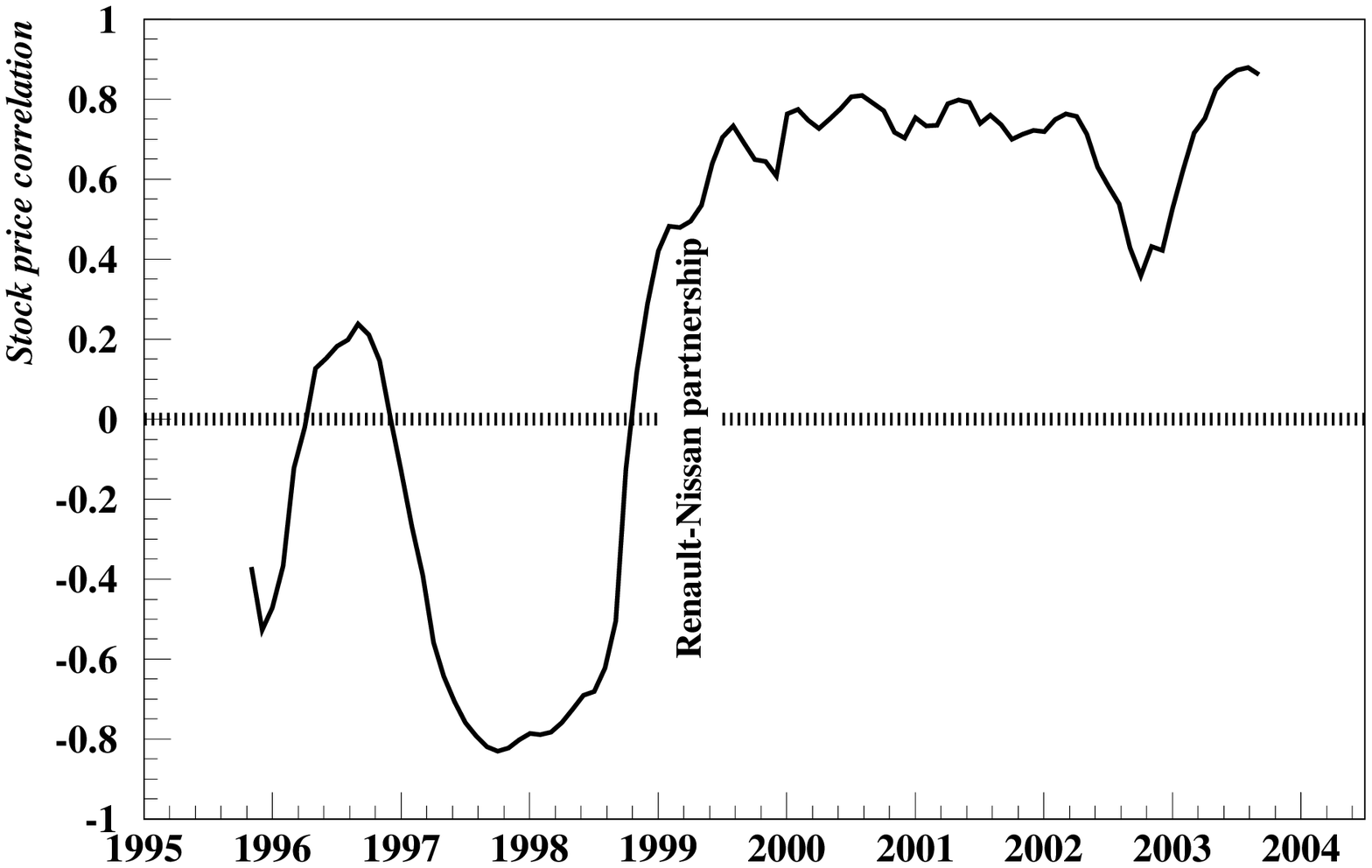}}
    {\bf Fig.1: Correlation between Nissan and Renault 
stock prices}.
{\small In April 1999, the French automaker took a 36.8\%
stake in Nissan; in March 2002, it increased it stake to 44.4\%
while at the same time Nissan acquired 15\% of Renault. The figure
shows that this cross-ownership brought about a marked increase
in the correlation between the stock prices of the two companies.
That this implication is not as obvious as could seem is illustrated by
the counterexample of Renault Argentina which, although 100\% 
owned by the parent company, shows little price correlation with
Renault itself.
The moving window used for computating the correlation
has a width of 21 months.
Assessing interaction strengths of public companies
is a key issue for a better understanding of stock markets and in
recent years a number of promising approaches have been tried
by several teams of econophysicists among which one
can mention Bonano et al (2001), 
Drozdz et al 2001, Kim et al (2004), 
Mantegna (1999), Menezes et al (2004),
Plerou et al (2001), Sornette et al (2003), Stauffer et al (1999).}
{\small \it Source: http://finance.yahoo.com/}.
 \end{figure}
%% --------------------------------------------------
\qpar

The fact that acquisitions, buyouts, takeovers and mergers are 
major corporate issues is attested by the warlike expressions
used in corporate parlance. Expressions such as blitzkrieg tender offer%
\qfoot{In a takeover, it is a tender offer which is so compelling that it
is accepted very quickly.}%
, 
scortched-earth policy, shark repellent, target company, porcupine
provision, sleeping beauty, and many others
were coined in order to describe
offensive and defensive tactics; they reveal that these questions
are major concerns of corporate management. 
\qpar

Cross-ownership and control by a majority owner would be
sufficient to substantially alter the traditional view, but this is only part
of the story. In the next section we analyze the effect of buyback programs.

%%%%%%% TABLE 

\begin{table}[htb]

 \small

\centerline{\bf Table 1\quad Repartition of the shares of DJI corporations} 

\vskip 3mm
\hrule
\vskip 0.5mm
\hrule
\vskip 2mm

$$ \matrix{
\hbox{Symbol} \hfill & \hbox{Corporation} \hfill &
\hbox{Shares} \hfill & \hbox{Treasury}  \hfill & 
\hbox{Shares} & \hbox{Non institutional} & \hbox{Price change}\cr
&&\hbox{authorized} \hfill & \hbox{stock}  & 
\hbox{oustanding} & \hbox{shares} & \hbox{9/10-9/17}\cr
&&\hbox{not yet issued} \hfill & \hbox{}  \hfill & 
\hbox{} & \hbox{} & \hbox{2001} \cr
\qtb
&&\hbox{} \hfill & \hbox{}  \hfill & 
\hbox{} & \hbox{[\%]} & \hbox{[\%]} \cr
\noalign{\hrule}
\qth
\hbox{AA}\hfill &\hbox{Alcoa}\hfill &
 &   6.4 &  100 & 19 & -11 \cr
\hbox{AIG}\hfill &\hbox{Am.Inter. Gr.}\hfill &
91&  5.4  & 100 &  32& -4.3 \cr
\hbox{BA}\hfill &\hbox{Boeing}\hfill &
42&  20  &  100&  14& -17 \cr
\hbox{C}\hfill &\hbox{Citigroup}\hfill &
&   5.9 &  100& 29 & -6.7 \cr
\hbox{CAT}\hfill &\hbox{Caterpillar}\hfill &
163&   19 &  100&  21& -7.1 \cr
\hbox{DD}\hfill &\hbox{Du Pont}\hfill &
80&   8.7 &  100&  37 &  -10\cr
\hbox{HD}\hfill &\hbox{Home Depot}\hfill &
75&   4.2 &  100&  32& -18 \cr
\hbox{HON}\hfill &\hbox{Honeywell}\hfill &
350&  6.3  &  100& 35 &  -10\cr
\hbox{IBM}\hfill &\hbox{IBM}\hfill &
132&   11 &  100& 11 & -17 \cr
\hbox{JNJ}\hfill &\hbox{Johns. and Johns.}\hfill &
177&   15  &  100& 36 & -3.2 \cr
\hbox{JPM}\hfill &\hbox{JP Morgan Ch.}\hfill &
45&  5.1  &  100&  36&  0.37\cr
\hbox{KO}\hfill &\hbox{Coca Cola}\hfill &
116&  0.30  &  100& 31 & -5.6 \cr
\hbox{MCD}\hfill &\hbox{McDonalds}\hfill &
178&  31  &  100&21  & -1.8 \cr
\hbox{MMM}\hfill &\hbox{3M}\hfill &
&  8.4  &  100&  24& -6.7 \cr
\hbox{MRK}\hfill &\hbox{Merk}\hfill &
143 &  34  &  100&  30& 1.7 \cr
\hbox{PFE}\hfill &\hbox{Pfizer}\hfill &
57&  14  &  100& 30 & 0.0 \cr
\hbox{SBC}\hfill &\hbox{SBC Comm.}\hfill &
&  3.9  &  100& 44 & 1.8 \cr
\hbox{UTX}\hfill &\hbox{United Techn.}\hfill &
290&   28 &  100& 14 & -26 \cr 
\hbox{XOM}\hfill &\hbox{Exxon Mobil}\hfill &
38&  22  &  100& 38 & -2.6 \cr
\hbox{}&\hbox{}&&    &  &  &  \cr
\qtb
\hbox{\bf Average}\hfill&\hbox{}&
\hbox{\bf 132}&  \hbox{\bf 15}  &  \hbox{\bf 100}&  
\hbox{\bf 28\%}& \hbox{\bf -7.2\%} \cr
\noalign{\hrule}
} $$
\vskip 1.5mm
Notes: The data refer to the situation in early 2004.
The numbers of shares outstanding have been normalized to
100. The number of shares issued is the sum of the treasury stock and
of the shares outstanding. 
The column labelled ``Non institutional shares'' gives the ratio:
(Shares issued not held by insiders or institutions or mutual funds)
/(shares issued). The fact that this ratio refers to the shares issued
rather than to the shares outstanding
reflects the fact that, once issued, shares can be moved out of or 
back into the treasury depending on the strategy of the company.
The table lists the companies for which adequate data could be
found in SEC 10-K reports.
Note that
we did not try to make a distinction between institutions such as 
investment banks on one hand
and mutual funds on the other hand,
because there is no clearly defined
borderline between them. For instance, State Street Corporation 
(NYSE: STT, market capitalization worth \$16 billions) is a financial 
holding company which is registered as an institution but through
its subsidiaries it has close links with the world of mutual funds as well.
\qL
Source: SEC Filings (10-K reports), http://finance.yahoo.com/
\vskip 2mm

\hrule
\vskip 0.5mm
\hrule
\vskip 3mm
\normalsize

\end{table}

\qI{Buyback programs}

Between 1995 and 1998, IBM bought back 772 millions of its shares%
\qfoot{WallStreetWishlist.com}%
,
which represents 46\% of its total shares oustanding (1690 million
shares in 2003). Over the same period of time, the price of its
stock jumped from \$15 in Januray 1995 to \$90 in December 1998.
Once bought back, these shares were withdrawn from the amount
of shares outstanding and kept in treasury stock (definitions of these
terms are recalled in Fig. 2). It is of interest to observe that
in March 2002, IBM's treasury stock held only 15\% of shares 
outstanding. 
What became of the rest of the shares which were bought back will be
considered in a moment.
\qpar

From Alcoa to Citigroup, to Merck, to Exxon Mobil, most
corporations implement similar buyback plans. For instance, between
1994 and 2003, Merck bought back 528 million shares which represents
24\% of its shares outstanding%
\qfoot{10-K Securities and Exchange Commission 
filing (March 10, 2004).}%
. 
Over the same period the price of its
stock jumped from \$10 to \$45. At global market level, announced 
buyback plans increased from \$26 billions in 1991 to \$236 billions in
2001. This later amount represents about 2.2\% of the annual trading 
volume on the New York Stock Exchange. 
\qpar
%%%% Fig.2
  \begin{figure}[htb]
    \centerline{\psfig{width=12cm,figure=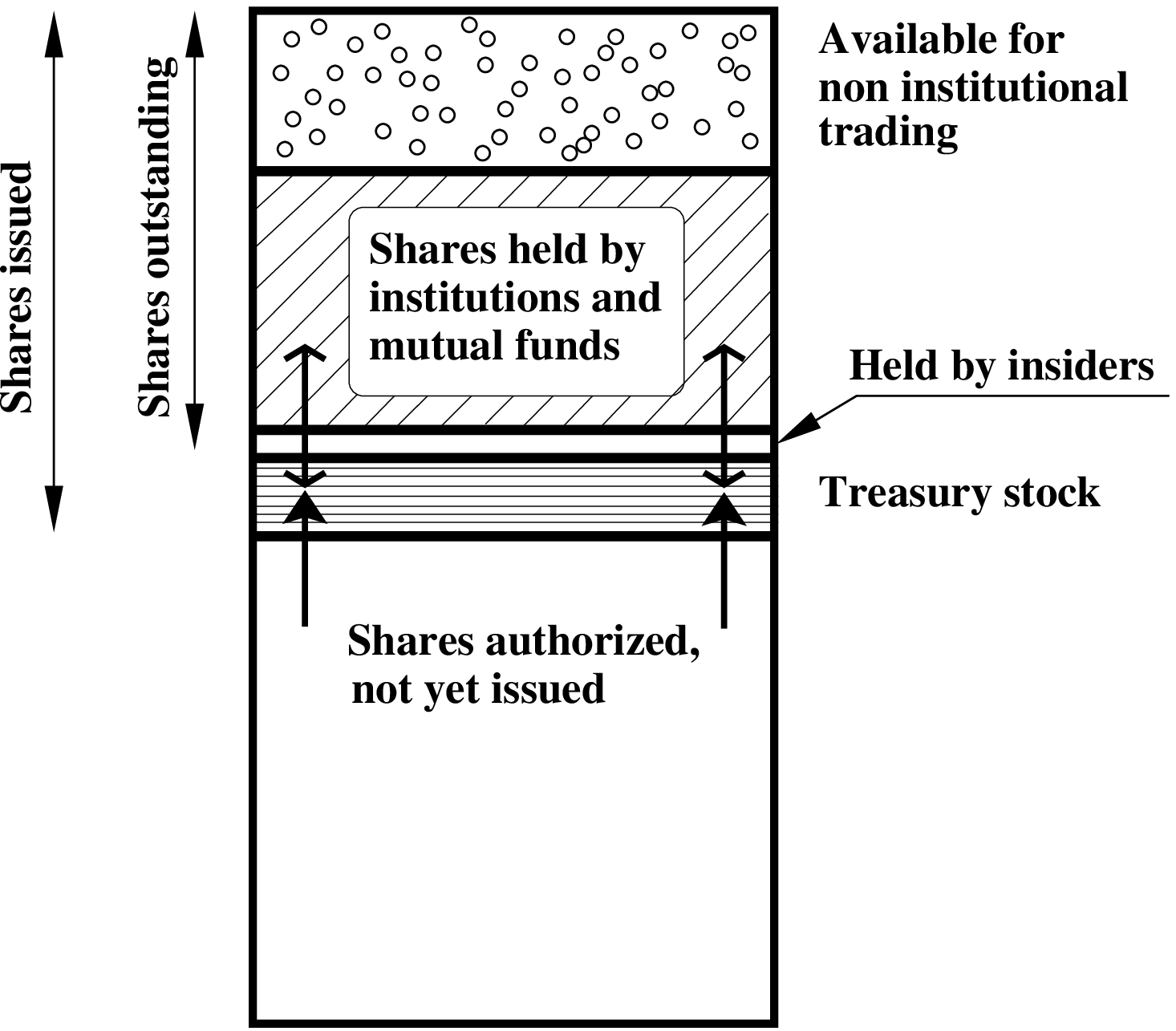}}
\qpar
    {\bf Fig.2: Schematic structure of the stock of large corporations.}
{\small Sizes of  the different components are
drawn in conformity with the averages given in table 1. 
As can be seen, the shares available for non-institutional
trading represent only the tip of the iceberg.
The mechanism by which shares authorized but not yet issued
are transformed into regular stock (up-pointing arrows)
amounts to a 
money-creation process.
During the period 1995 to 2000, on the basis of an average  NYSE
capitalization of the order of \$9,700 billions, the
potential of this mechanism amounted to three times the
increase of the M3 stock of money in the United States.
Such a potential is
of particular importance in takeover operations.}
{\small \it Source: Table 1}.
 \end{figure}
%% --------------------------------------------------
\qpar
The treasury stock has become an important component in the
strategy of many corporations. This is shown by its mere magnitude.
From the data extracted from the 10-K reports destined to the
Securities and Exchange Commission, it can be seen (table 1) that
the treasury stock represents on average 15\% of the shares
outstanding. The three largest percentages are Coca Cola (43\%),
Merck (34\%) and McDonalds (31\%), while the three smallest
are JP Morgan Chase (0.30\%), SBC Communications (3.9\%)
and Walt Disney (4.2\%). 
\qpar

With the development of stock options, the treasury stock has become
even more important than it used to be. When an executive exercises
her options by buying 100,000 shares where do these shares come from?
Usually, from the treasury stock. Naturally, to generate a profit
the shares  must be resold immediately at market price
after having been bought at the reduced option price. 
The net result of the operation is that the shares have been moved
from the treasury to the set of outstanding shares. Repeated many times,
such operations tend to inflate the amount of oustanding shares,
something stockholders see with much displeasure because it lessens
the dividends to be paid for each share%
\qfoot{At this point, one should mention the fact that over the last
decade more and more companies gave up distributing dividends; 
other ways for rewarding shareholders have been introduced among
which share buyback is the most important.}%
.
This leads the Boards of Directors to set up buyback plans%
\qfoot{This is announced in the SEC filings by sentences like the
following: ``IBM will repurchase shares on the open market or in
private transactions'' (May 7, 2004).}%
.
\qpar

Of course, there may be other reasons as well such as for instance 
to support stock prices.
In the repurchase of shares
on the open market, SEC regulation imposes that on any single
day, the purchases do not exceed 25\% of the average daily trading
volume. However, the SEC has the ability to loosen these restrictions
in special circumstances for instance after a crash. It did so in the
days following September 11, 2001 (Gabelli 2003). The following
paragraph provides a test of the role of treasury stock in
this connection.
\qpar

{\bf Role of treasury stock in the days after 9/11}\quad The last 
column of table 1 gives the price change between September 10, 2001
and September 17 when the market reopened. If we discard Boeing 
and United Technologies which are obvious outliers due to their links
with the aerospace industry (which was especially hard hit), there is
a correlation of 0.44 between sizes of treasury stock and price changes.
In other words, the larger the treasury stock, the smaller was the
price fall. The three companies with the smallest treasury 
(2.8\% on average), namely
JP Morgan Chase, SBC Communications and Walt Disney
experienced a decrease of 7.3\%, whereas at the other end
of the spectrum, companies such as 
Coca Cola, Merck and MacDonalds with an average treasury of 36\%
experienced a price {\it increase} of 0.11\%.
This does not necessarily mean that corporations stepped in massively
to support their stock; it can also be interpreted by saying that knowing
that, if necessary (that is to say in case of a substantial drop)
they {\it could do so}, was certainly reassuring for institutional investors
who therefore abstained from selling. The effect shown by 9/11 could be
confirmed by studying other crash episodes. More on this will be said in 
a subsequent paper.
\qpar

Another important usage of treasury stock is for acquisitions.
For instance, when Alcoa acquired part of Aluminio in 2002, it used
17 million shares representing 2.1\% of its shares outstanding.
Similarly, back in 2001, when the American International Group
bought American General Corporation in a major takeover, it used
\$2.3 billions of its treasury stock. As can be seen in table 1,
in most cases the amount of shares authorized greatly exceeds 
the total amount of shares oustanding plus treasury shares.
This gives companies a kind of money-creation power: by issuing
new shares, the treasury stock can (under some conditions) be
replenished and used for acquisitions and stock options operations.

\qI{Merger and acquisitions}

In order for a company $ A $ to purchase some or all outstanding shares of
a company $ B $, there are basically three procedures. 
\qbu {\it Open market purchases} at the current market price.
\qbu {\it Negociated private transaction}. In this strategy, company $ A $ will
try to convince a big investor to sell a large block of stock.
\qbu {\it Tender offer}. In this case, company $ A $ offers to repurchase
a fixed amount of stock at a specified price within a period which is usually
of the order of one month. In order to make the offer attractive, the price
is set at level above the current market price. Very often the announcement
of a tender offer brings about a sudden and huge price increase.
\qpar

For instance, in a recent case, after Omnicare (NYSE: OCR) announced
a tender offer to purchase all of the outstanding shares of Neighborcare
(NASDAQ: NCRX, market capitalization in June 2004 worth \$1.4 billion),
the price of Neighborcare stock jumped from \$17 on May 20, 2004 to
\$29 on May 25, 2004, a 70\% increase within 3 business days. 
\qpar

Incidentally, such an effect which 
is not only completely deterministic but also fairly predictable,
does not fit well with the random walk hypothesis. Initial Public Offerings
(IPO) are another example of price evolution shaped by
underwriters. Indeed, the actual offering is deliberately set at a level
which will create an artificial shortage and send the price upward.
But that price increase will be followed by a dip when the company 
insiders sell their shares. Indeed,
according to regulation they must wait
at least 90 days (the so-called lock up period) before they can sell, which
means that the price is bound to drop about 3 months after the initial 
offering. 
\qpar

Coming back to the merger and acquisition procedure, we see that 
whatever strategy is used, it tends to push up the price of the shares
of company $ B $. One may therefore expect intense
merger and acquisition activity to produce an overall increase of
stock price levels. The 1990s were indeed a time of high merger 
activity marked 
in addition by several multibillion mega-mergers, for instance
Merk/Medco (1993), 
Exxon/Mobil (1998), BP/Amoco/Arco (1998-1999), Total/Elf/Fina (1998-1999),
Chevron/Texaco/\-Caltex (2001). 
Over the period 1991-2000 there were on average
5,500 mergers and acquisitions annually as compared to 2,500 during
the previous decade. As shown in Fig.3a, there was a fairly close 
correlation between the rise in merger and acquisition activity and
the increase in stock prices (the correlation is equal to 0.95). If
mergers tend to push up stock prices, in return inflated stock prices
facilitate takeovers because usually
a substantial fraction of the deal is paid in shares.
Because of this two-way interdependence, it would
be pointless to ask which was the cause and which
the consequence. 
\qpar
%%%% Fig.3
  \begin{figure}[htb]
    \centerline{\psfig{width=15cm,figure=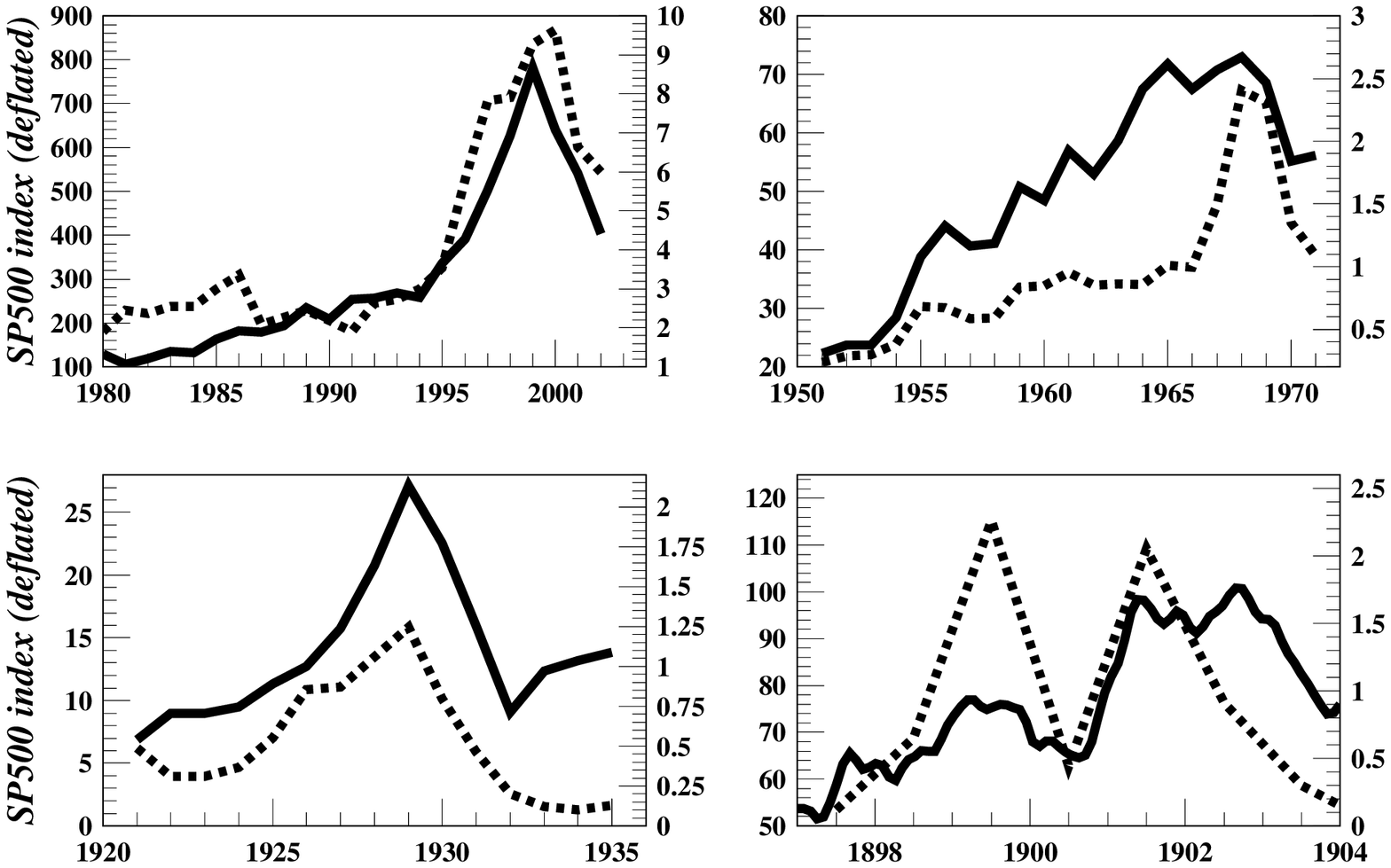}}
    {\bf Fig.3: Relationship between merger and acquisition activity
on one hand and overall level of stock prices on the other hand.}
{\small Solid line: stock price level (left-hand scale); broken line:
number of mergers and acquisitions in thousands (right-hand scale).
Four major price peaks are considered. (a) Upper left-hand panel:
price increase which started in 1980 and gathered speed in the
1990s; the regression coefficient of percentage changes is $ 0.27 $.
(b) Upper right-hand panel: the price increase started in the 1950s and
culminated in 1968; the regression coefficient of percentage 
changes is $ 0.32 $. (c) The lower left-hand panel
shows the price peak which
led to the crash of october 1929; the regression coefficient of percentage 
changes is $ 0.29 $. (d) Lower right-hand panel: the price peak which
culminated in 1902 was substantially smaller and shorter than 
subsequent ones; the small number of data points does not permit
to compute regression coefficients in a reliable way. 
Purchasing a company is an operation which boosts the demand
for stocks and pushes up their price level. In return higher stock prices
facilitate takeovers because a sizeable part of the deal may be
paid in shares instead of cash. As a result this is a two-way 
interaction and it would be meaningless to wonder which of the
two factors is the cause and which the consequence. }
{\small \it Sources: Historical Statistics of the United States (1975),
Farrel (1972), http://finance.yahoo.com}.
 \end{figure}
%% --------------------------------------------------
Fig. 3b,c,d shows that the same scenario repeated itself during former
price peaks. The price peak of 1897-1904 was much shorter than
subsequent price peaks. The small number of points did not allow
to compute regression estimates, but for the other three cases the
regression coefficients of percentage changes are fairly close around
0.30. On average, we get the following relationship:
$$ { \Delta \ \hbox{Stock prices} \over \hbox{Stock prices} } =
a{ \Delta\ \hbox{Mergers} \over \hbox{Mergers} }+b \quad
\hbox{where: } a=0.29\pm 0.14,\quad b=5.0\pm 4 $$

\qI{Conclusion}

In the face of the above evidence, why are researchers reluctant to
incorporate corporations as direct players into their multiagent, multifractal
models? Why do economists continue to discuss the question of
whether or not there are indeed speculative peaks without taking into
account the fact that buyback programs added to merger and acquisition
activity force upon the market an articial dearth of stocks which 
invalidates value-based expectations? There is probably a historical 
reason to this.
\qpar

For a long time, stock markets have indeed been working as fairly
competitive markets. There were no stock options, 
fewer public companies were controlled by large corporations,
merger and acquisitions had to be paid for mostly in cash. 
The creation of derivative markets was a major change that 
occurred over the last 20 years. Because of their novelty, these
markets required new theoretical tools which have indeed been
developed (see the vast literature on
Black-Scholes Option Pricing Models). However, probably because
structural novelties in stock markets were less appearant, 
the major changes
described in this paper failed to be incorporated into
standard models. 
\qpar

This study continues previous work (Roehner 2001, 2002, 
Maslov and Roehner 2004) in which we called into question some of the
underlying assumptions of standard models. Once
the leading role of corporations is recognized and accepted, 
one of the first tasks will be to identify and
estimate the cross-ownership connections as we did
for Renault and Nissan. 
Because, stock markets
are multi-faceted, many additional studies will be required
in order to work out and formalize the present framework and to
derive all its implications. 
\qpar

{\bf Acknowledgment}\quad I am most grateful to Professor 
Bruce Mizrach (Rutgers) for sharing with me his thorough 
knowledge of stock market mechanisms. 

\vfill \eject

{\large \bf References}
\vskip 5mm

\qparr
Bonanno (G.), Lillo (F.), Mantegna (R.) 2001: Levels of complexity
in financial markets. Physica A 299,16-27.

\qparr
Drozdz (S.), Kwapien (J.), Gr\"ummer, Ruf (F.), Speth (J.) 2001:
Quantifying the dynamics of financial correlations.
Physica A 299,144-153.

\qparr
Farrel (M.L.) 1972: The Dow Jones averages 1885-1970.
Dow Jones Books. Princeton.

\qparr
Gabelli and Company 2003: Tender offer. The Dutch variety. 
(http://www.gabelli.com, April 21, 2003).

\qparr
Kim (K.), Yoon (S.-M.), Choi (J.S.), Takayasu (H.) 2004: Herd behaviors
in financial markets.
Preprint available on: http://arXiv.org/abs/cond-mat/0405172 (9 May).

\qparr
Mantegna (R.N.) 1999: Hierarchical structure in financial markets.
The European Physical Journal B 11,193-197.

\qparr 
Maslov (S.), Roehner (B.M.) 2004: The conundrum of stock versus 
bond prices.
Physica A 335,164-182.

\qparr
Menezes (M.A. de), Barab\'asi (A.-L.) 2004: Separating internal and
external dynamics of complex systems. 
Preprint available on: http://arXiv.org/abs/cond-mat/0406421 (18 June).

\qparr
Plerou (V.), Gopikrishnan (P.), Rosenow (B.), Amaral (L.A.N.),
Stanley (H.E.) 2001: Collective behavior of stock price movements:
A random matrix theory approach. 
Physica 299,175-180.

\qparr
Prigge (S.) 1998: A survey of German corporate governance. 
in Hopt (K.J.), Kanda (H.), Roe (M.J.), Wymeer (E.), Prigge (S.) eds,
Claredon Press, Oxford.

\qparr
Roehner (B.M.) 2001: Hidden collective factors in speculative trading.
Springer-Verlag. Berlin.

\qparr
Roehner (B.M.) 2002: Patterns of speculation. 
Cambridge University Press, Cambridge.

\qparr
Sornette (D.), Gilbert (T.), Helmstetter (A.), Ageon (Y.) 2003: 
Endogenous versus
exogenous shocks in complex networks: an empirical test.
Preprint available on: http://arXiv.org/abs/cond-mat\-/0310135. 

\qparr
Stauffer (D.), Sornette (D.) 1999: Self-organized percolation model
for stock market fluctuations. 
Physica A 271, 496-506.

\end{document}